\documentclass[preprintnumbers, prd, twocolumn, showpacs, floatfix, preprintnumbers, 
letterpaper, 
superscriptaddress,nofootinbib]{revtex4-1}
\pdfoutput=1
\usepackage{graphicx,epsfig}
\usepackage{amsmath,mathrsfs,amsfonts}
\usepackage {amssymb}
\usepackage {longtable}
\usepackage{multirow}
 \usepackage{enumerate}
 
\usepackage{dcolumn}
\usepackage{bm}
\usepackage{amsfonts}
\usepackage{subfigure}
\usepackage{color}
\usepackage{relsize}

\newcommand{\be}{\begin{equation}}
\newcommand{\ee}{\end{equation}}
\newcommand{\bea}{\begin{eqnarray}}
\newcommand{\eea}{\end{eqnarray}}

\newcommand{\abs}[1]{\lvert#1\rvert}

\begin{document}

%\title{Scaling attractors in the full Israel-Stewart formalism for viscous cosmologies}  

\title{Dynamics of viscous cosmologies in the full Israel-Stewart formalism}

\author{Samuel Lepe}
\email{samuel.lepe@pucv.cl}
\affiliation{Instituto de F\'{\i}sica, Pontificia Universidad Cat\'olica de 
Valpara\'{\i}so, 
Casilla 4950, Valpara\'{\i}so, Chile}

\author{Giovanni Otalora}
\email{giovanni.otalora@pucv.cl}
\affiliation{Instituto de F\'{\i}sica, Pontificia Universidad Cat\'olica de 
Valpara\'{\i}so, 
Casilla 4950, Valpara\'{\i}so, Chile}

\author{Joel Saavedra}
\email{joel.saavedra@pucv.cl }
\affiliation{Instituto de F\'{\i}sica, Pontificia Universidad Cat\'olica de 
Valpara\'{\i}so, 
Casilla 4950, Valpara\'{\i}so, Chile}

\date{\today}

\begin{abstract} 
It is performed a detailed dynamical analysis for a bulk viscosity model in the full Israel-Stewart formalism for a spatially flat Friedmann-Robertson-Walker Universe. In our study we have considered the total cosmic fluid constituted by radiation, dark matter and dark energy. The dark matter fluid is treated as an imperfect fluid which has a bulk viscosity that depends on its energy density in the usual form $\xi(\rho_{m})=\xi_{0} \rho_{m}^{1/2}$, whereas the other components are assumed to behave as perfect fluids with constant EoS parameter. We show that the thermal history of the Universe is reproduced provided that the viscous coefficient satisfies the condition $\xi_{0}\ll 1$, either for a zero or a suitable nonzero coupling between dark energy and viscous dark matter. In this case, the final attractor is a dark-energy-dominated, accelerating Universe, with effective EoS parameter in the quintessence-like, cosmological constant-like or phantom-like regime, in agreement with observations. As our main result, we show that in order to obtain a viable cosmological evolution and at the same time alleviating the cosmological coincidence problem via the mechanism of scaling solution, an explicit interaction between dark energy and viscous dark matter seems inevitable. This result is consistent with the well-known fact that models where dark matter and dark energy interact with each other have been proposed to solve the coincidence problem. Furthermore, by insisting in above, we show that in the present context a phantom nature of this interacting dark energy fluid is also favored.

\end{abstract}


\maketitle

%%%%%%%%%%%%%%%%%%%%%%%%%%%%%%%%%%%%%%%%%%%%%%%%%%%%%%%%%%%%%%%%%%%%
\section{Introduction}\label{Introduction}
%%%%%%%%%%%%%%%%%%%%%%%%%%%%%%%%%%%%%%%%%%%%%%%%%%%%%%%%%%%%%%%%%%%%

Dark energy and dark matter are the most important energy components which dominate the current dynamics of the Universe.  The dark energy component can be described as a cosmic fluid characterized by a negative pressure which is responsible for the current cosmic acceleration and it constitutes the $70\%$ of the total energy of the present Universe \cite{Riess:1998cb, Perlmutter:1998np,Ade:2013zuv}. On the other hand, the dark matter component is a pressureless matter fluid which interacts only gravitationally \cite{Zwicky:1933gu}. Its energy fraction is about $25\%$, and ordinary (baryonic) matter and radiation is only about $5 \%$. Unlike dark energy,  dark matter can cluster by gravitational instability and it has played a crucial role for the growth of large-scale structure such as galaxies and clusters of galaxies \cite{AmendolaTsujikawa}.

A description based in fluids is thus a natural setting in studying the dynamics of the various energy components of the Universe \cite{Copeland:2006wr, AmendolaTsujikawa}. In the simplest picture, these energy components are assumed to be barotropic perfect fluids which are completely  characterized by the energy density $\rho$,  pressure density $p$ and equation of state (EoS) parameter $w=p/\rho$, with $w$ equal to a constant. For instance, in the case of radiation the EoS parameter takes the value $w_{r}=1/3$, whereas for dark matter and baryons the EoS parameter is $w_{m}=0$, which is adequate for nonrelativistic matter. Also, since there is no sufficient evidence for or against an evolving dark energy component \cite{Copeland:2006wr,F(R)}, we can assume a dark energy model with constant EoS parameter, which satisfies the constraint $w_{DE}<0$, including the case of a cosmological constant with $w_{DE}=-1$ \cite{AmendolaTsujikawa}. 

Nevertheless, from the viewpoint of this fluid description, and the current cosmological observational data, there is no reason for excluding a more real setting where these cosmic fluids are imperfect fluids due to a bulk viscosity in them \cite{Foot:2014uba,Fan:2013tia,Bamba:2012cp}.
For instance, in an expanding Friedmann-Robertson-Walker (FRW) Universe, with $H=\dot{a}/a$ the Hubble expansion rate, the usual energy conservation equation for a barotropic perfect fluid is given by $\dot{\rho}+3 H\gamma\rho=0$, where we have defined the useful parameter $\gamma\equiv 1+w$. Now, if dissipative processes occur due to a bulk viscosity in the fluid, its energy conservation equation is modified in the following form 
\be
\dot{\rho}+3 H\left[\gamma\rho+\Pi\right]=0, 
\label{ModConsEq}
\ee where the additional term $\Pi$ entering in the brackets represents the viscous pressure of the fluid. In this more real setting the nature of $\Pi$ is determined by the approach or formalism adopted in the description of the viscous fluid.

In general there are two classes of formalisms that are used in the description of viscous fluids in Cosmology.  One of them is the noncausal Eckart formalism \cite{Eckart:1940zz} where the usual choice for the viscous pressure is $\Pi=-3 \xi(\rho) H$, and $\xi(\rho)$ is the bulk viscosity coefficient which is a function of the energy density $\rho$. This formalism has been applied by many authors in the study of viscous cosmologies in a flat FRW Universe, in both the background \cite{Velten:2013qna, Leyva:2016gzm, Cruz:2014iva, Avelino:2013wea}, as well as in the perturbation level \cite{Velten:2014xca,Velten:2013pra,Velten:2013rra,Velten:2012uv}. The other one is the causal Israel-Stewart (IS) formalism \cite{Israel:1979wp} in which the viscous pressure $\Pi$ satisfies the transport equation
\be
\tau \dot{\Pi}+\left(1+\frac{1}{2}\tau \Delta\right)\Pi=-3 H \xi(\rho), 
\label{TransEq}
\ee where $\Delta\equiv 3 H-\frac{\dot{\tau}}{\tau}-\frac{\dot{\xi}}{\xi}-\frac{\dot{T}}{T}$, with $T$ being the barotropic temperature of the fluid, and $\tau$ is the relaxation time, which is of the order of the mean interaction time \cite{Maartens:1996vi}. For very small values of the relaxation time $\tau$, as occurs for example in the early Universe where we have only short distances, the two formalisms are indistinguishable and thus the use of the noncausal Eckart formalism is completely justified. However, at late times this noncausal approach does not work, and it is then reasonable to opt by the causal IS formalism \cite{Cruz:2016rqi,Cruz:2017lbu}. 
In order to calculate $T$, the Gibbs integrability condition is used, and for a barotropic fluid with constant EoS parameter, it takes the form $T\sim \rho^{\left(\gamma-1\right)/\gamma}$ \cite{Maartens:1996vi}. On the other hand, to calculate the relaxation time $\tau$, the relation between $\tau$ and the speed of bulk viscous perturbations $c_{b}$ is also used. In the case of a barotropic fluid with constant EoS parameter, it takes the form $\tau=\frac{1}{\left(2-\gamma\right)\gamma}\frac{\xi(\rho)}{\rho}$ \cite{Cruz:2016rqi}. Finally, both in the noncausal, as well as in the causal formalism, the usual choice for the bulk viscosity function is a power-law of the energy density, $\xi(\rho)=\xi_{0}\rho^{s}$, where $\xi_{0}$ is a positive constant to be estimated from the comparison with cosmological observations, and $s$ is an exponential parameter \cite{Cruz:2016rqi,Cruz:2014iva}. 

In principle, any one of them, radiation, dark matter or dark energy, could be a viscous fluid, but since dark matter can cluster whereas dark energy does not cluster, and it has provided the most important contribution to the formation of structures in the Universe, it is then interesting to study the physical viability of a viscous dark matter fluid in an expanding FRW Universe. Furthermore, since does not exist any physical principle that excludes a possible explicit coupling between dark energy and dark matter, then we have the freedom in considering an even more general setting where this viscous dark matter fluid is explicitly interacting with dark energy \cite{Amendola:1999er, Guo:2007zk,Zimdahl:2001ar,CalderaCabral:2008bx, Gumjudpai:2005ry,Avelino:2013wea}. In this way, in the present paper we use the perspective of dynamical systems to investigate the cosmological dynamics of a nonevolving dark energy fluid which is explicitly interacting with a viscous dark matter fluid in the framework of the causal IS formalism. In particular, here we deepen the study of ``Scaling Solutions''. The study of dynamical systems plays an important role in Cosmology, since it allows to check the cosmological viability of a model, in reproducing the thermal history of the Universe and the current cosmic acceleration \cite{Coley:2003mj}. Furthermore, the mechanism of scaling solutions has a very relevant place in this apparatus. These are attractors in which the dark energy density mimics the background matter energy density and thereby providing a new line of attack on the fine-tuning problem, or cosmological coincidence problem of dark energy \cite{Amendola:1999qq,Dutta:2016dnt,Wang:2016lxa,Zimdahl:2002zb,Quartin:2008px}.

The organization of the paper is as follows: in Sec. \ref{Dyncs} we present the field equations for a flat FRW Universe filled with radiation, viscous dark matter and dark energy, obeying a barotropic EoS parameter all them. In this section we also introduce the phase space variables, and then we rewrite the field equations in terms of them.  In Sec. \ref{DyncsA} we study the dynamics of this model in the case of a zero coupling between dark energy and viscous dark matter, whereas in Sec. \ref{DyncsB} it is studied the case of a nonzero coupling. Finally, Sec. \ref{Conclusions} is devoted to conclusions. Throughout the paper, we adopt natural units and $8\pi G=1$.

%%%%%%%%%%%%%%%%%%%%%%%%%%%%%%%%%%%%%%%%%%%%%%%%%%%%%%%%%%%%%%%%%%%%

\section{Cosmological dynamics in the Israel-Stewart (IS) formalism }\label{Dyncs}
%%%%%%%%%%%%%%%%%%%%%%%%%%%%%%%%%%%%%%%%%%%%%%%%%%%%%%%%%%%%%%%%%%%%
In applying the causal IS formalism to a cosmic viscous fluid in a realistic Universe, we consider that the total cosmic fluid is constituted by radiation, dark matter and dark energy. Moreover, according to the standard cosmological model, the formation of structure of the Universe (galaxies, clusters) had its most important development during the epoch in that dark matter was the dominant component \cite{AmendolaTsujikawa}. It is then interesting to investigate the physical viability and cosmological consequences of any modification to the dark matter dominated era. In the following we assume that the dark matter fluid is a viscous fluid which responds to the causal IS formalism. So, the Friedmann constraints and the conservation equations for radiation, dark matter and dark energy components, are written as
\bea
\label{Fr00}
&& 3 H^2=\rho_{r}+\rho_{m}+\rho_{DE},\\ \label{CEqr}
&& \dot{\rho}_{r}=-4 H \rho_{r},\\ \label{CEqm}
&& \dot{\rho}_{m}=-3 H \rho_{m}-3 H \Pi+Q,\\ \label{CEqDE}
&& \dot{\rho}_{DE}=-3 H \gamma_{DE}\rho_{DE}-Q,
\eea where total energy density is $\rho_{T}=\rho_{r}+\rho_{m}+\rho_{DE}$ with $\rho_{r}$, $\rho_{m}$ and $\rho_{DE}$ the energy densities of radiation, dark matter and dark energy, respectively.
The EoS parameter for dark energy is expressed in terms of the barotropic parameter $\gamma_{DE}\equiv 1+w_{DE}<1$, whereas for radiation and dark matter we have $\gamma_{r}\equiv 1+w_{r}=4/3$ and $\gamma_{m}\equiv 1+w_{m}=1$, respectively. The dark matter fluid is a viscous fluid with $\Pi$ satisfying the transport equation \eqref{TransEq} in the IS formalism, and $Q$ represents a possible coupling between viscous dark matter and dark energy. Considering a coupling between dark matter and dark energy is reasonable since there is no physical principle that excludes an explicit interaction between them \cite{Copeland:2006wr,Amendola:1999er}. The total energy density $\rho_{T}$ satisfies the conservation equation $\dot{\rho}_{T}+3 H\gamma_{eff}\rho_{T}=0$, where $\gamma_{eff}\equiv 1+w_{eff}$ and the effective EoS parameter takes the value  $w_{eff}=p_{eff}/\rho_{T}=(p_{T}+\Pi)/\rho_{T}$, such that $p_{T}=p_{r}+p_{m}+p_{DE}$ is the total pressure density from the  energy components. Thus, by taking in account the negative pressure from the bulk viscosity, the accelerated expansion occurs for $w_{eff}<-1/3$ or equivalently for $\gamma_{eff}<2/3$. As usual, we take the bulk viscous function $\xi(\rho_{m})$ to be a power-law of the energy density of dark matter, in the form $\xi(\rho_{m})=\xi_{0}\rho_{m}^{1/2}$, with  $\xi_{0}$ a positive parameter. Here, we have assumed that the exponential parameter $s$ is equal to $1/2$ because it is the simplest choice for which we can rewrite the cosmological equations in the form of an autonomous system \cite{Cruz:2014iva}.

To study the dynamics of the model, we introduce the following set of dimensionless variables 
\be
x=\Omega_{DE}=\frac{\rho_{DE}}{3 H^2},\:\:\:\: y=\Omega_{m}=\frac{\rho_{m}}{3 H^2},\:\:\:\: z=\frac{\Pi}{3 H^2}.
\label{PhaseSpaceV}
\ee In terms of these variables we have that
\be
w_{eff}=\gamma_{eff}-1=\left(\gamma_{DE}-\frac{4}{3}\right)x+z-\frac{y}{3}+\frac{1}{3}.
\ee  Also, we rewrite the set of cosmological equations  \eqref{Fr00},\eqref{CEqr}, \eqref{CEqm} and \eqref{CEqDE}, along with the transport equation \eqref{TransEq},  in the following form
\bea
&&\frac{dx}{dN}=x\left( 3 x{{\gamma}_{{DE}}}-3 {{\gamma}_{DE}}+3 z-y-4 x+4\right)-\frac{Q}{3 H^3},\nonumber \\
&& \frac{dy}{dN}=3 x y {{\gamma}_{DE}}+\left( 3 y-3\right) z-{{y}^{2}}+\left( 1-4 x\right) y +\frac{Q}{3 H^3},\nonumber \\
&& \frac{dz}{dN}=3 z \left(x {{\gamma}_{DE}}+ {{z}}\right)-z \left(\frac{3 z}{2 y}+\frac{\sqrt{3 y}}{\xi_{0}}\right)+\nonumber\\
&& \left( 1-4 x\right) z-y \left(z+3\right) +\frac{Q}{6 {{H}^{3}}}\frac{z}{y},
\label{AutoSys} 
\eea where we have introduced the e-folding number $N=\log(a)$ which is convenient to use for the
dynamics of dark energy \cite{Copeland:2006wr}. This set of three first-order differential equations will be an autonomous system for the three independent variables $x$, $y$, and $z$, if the coupling $Q$ can be written in terms of them. In this case, the critical points $(x_{c},y_{c},z_{c})$ of the above dynamical system can be extracted by imposing the conditions $\frac{dx}{dN}=\frac{dy}{dN}=\frac{dz}{dN}=0$. Finally, perturbing the system linearly around these critical points, and expressing the perturbations equations in terms of a perturbation matrix, allows one to determine the type and stability of each  critical point by examining the eigenvalues of this matrix  \cite{Coley:2003mj,Copeland:2006wr}. In a scenario cosmologically viable, the complete thermal history of the Universe must be reproduced. In early times, after end of inflation the Universe is dominated by radiation, followed by a dark matter-dominated era around the redshift $z=3000$, and finally it becomes dominated by dark energy in late-times around the redshift $z\sim 1$. 

Several different forms of the coupling between dark energy and dark matter have been proposed in the literature. One possibility usually studied in scalar field models is to consider an interaction of the form $Q\sim \rho_{m} \dot{\phi}$ with $\dot{\phi}$ the velocity of the homogeneous scalar field, see Refs. \cite{Copeland:2006wr,Gumjudpai:2005ry,Amendola:1999er}. A second approach more close to the description of fluids consists in introducing an interaction in the form $Q\sim \Gamma \rho_{m}$ or  $Q\sim \Gamma \rho_{DE}$,  with the normalization of the factor $\Gamma$ in terms of the Hubble parameter $H$, i.e. $\Gamma/H =\alpha$ , where $\alpha$ is a dimensionless constant \cite{AmendolaTsujikawa}. In this way, in the following we study the corresponding dynamical system for two different cases of the coupling $Q$. In the first case, we study the dynamical system corresponding to $Q=0$.  In the second case, we consider the simplest nonzero coupling in the form  $Q=\alpha H \rho_{DE}$, with $\alpha$ the coupling constant. We consider this coupling since another coupling $Q\sim H \rho_{m}$ excludes the solution $y_{c}=\Omega_{m}=1$ (dark matter-dominated era), as it can be verified from Eqs. \eqref{AutoSys}.

\def\tablename{Table}%
\begin{table*}[ht]
\centering
\begin{center}
\begin{tabular}{c c c c c c c c}\hline\hline
Name &  $x_{c}$ & $y_{c}$ & $z_{c}$ &  Existence & $\gamma_{eff}$&  Acceleration & 
Stability 
\\\hline 
\\ 
$R_{1}$ & $0$ & $0$ & $0$ & Always & $\frac{4}{3}$  & Never  & Saddle point for \\&&&&&&&  all values  \\ \\ 
$R_{2-}$ & $0$ & $1$ & $\frac{1-\sqrt{6 {{{\xi}_{0}}^{2}}+1}}{\sqrt{3} {{\xi}_{0}}}$ & Always  &  $1+\frac{1-\sqrt{6 {{{\xi}_{0}}^{2}}+1}}{\sqrt{3} {{\xi}_{0}}}$  & $\xi_{0}>\frac{2\sqrt{3}}{17}$ & Stable node for \\&&&&&&&  $\xi_{0}>\xi_{0}^{*}(\gamma_{DE})$  with \\&&&&&&&  $1-\sqrt{2}<\gamma_{DE}<1$ \\ &&&&&&&
Saddle point for \\&&&&&&& $\xi_{0}<\xi_{0}^{*}(\gamma_{DE})$ with \\&&&&&&&  $1-\sqrt{2}<\gamma_{DE}<1$ or \\ &&&&&&&
for $\gamma_{DE}<1-\sqrt{2}$ \\ \\
$R_{2 +}$ 
& $0$ & $1$ & $\frac{1+\sqrt{6{{{\xi}_{0}}^{2}}+1}}{\sqrt{3}{{\xi}_{0}}}$ &  Always & $1+\frac{1+\sqrt{6{{{\xi}_{0}}^{2}}+1}}{\sqrt{3}{{\xi}_{0}}}$& Never & Unstable node for \\&&&&&&&  all values\\ \\
$R_{3}$ 
& $1-\frac{z_{c}}{{{\gamma}_{DE}}-1}$ & $\frac{z_{c}}{{{\gamma}_{DE}}-1}$ & $\frac{3{{{\xi}_{0}}^{2}}{{\left( {{{\gamma}_{DE}}^{2}}-2{{\gamma}_{DE}}-1\right)}^{2}}}{4\left( {{\gamma}_{DE}}-1\right)}$ &  $\gamma_{DE}-1<z_{c}<0$ with \\&&&&  $1-\sqrt{2}<\gamma_{DE}<1$ & ${{\gamma}_{DE}}$  & $\gamma_{DE}<\frac{2}{3}$ &  Stable node for \\&&&&&&&  $\xi_{0}<\xi_{0}^{*}(\gamma_{DE})$ with \\&&&&&&&  $1-\sqrt{2}<\gamma_{DE}<1$\\ &&&&&&& Saddle point for \\&&&&&&&  $\xi_{0}>\xi_{0}^{*}(\gamma_{DE})$ 
with \\&&&&&&& $1-\sqrt{2}<\gamma_{DE}<1$ \\\\
$R_{4}$ 
&  $1$ & $0$ & $0$    & Always  &  ${{\gamma}_{DE}}$  & $\gamma_{DE}<\frac{2}{3}$ & Stable node for \\&&&&&&& 
 $-1<\gamma_{DE}<1-\sqrt{2}$ \\&&&&&&& Stable spiral for \\&&&&&&&  $\gamma_{DE}<-1$ \\&&&&&&& Saddle point for 
\\&&&&&&& 
$1-\sqrt{2}<\gamma_{DE}<1$\\
\hline\hline
\end{tabular}
\end{center}
\caption{The physical critical points of the dynamical system  \eqref{AutoSyszeroQ} with coupling $Q=0$, and their existence and stability conditions, along with the  conditions for 
accelerated expansion, and the corresponding values of the effective equation of state $\gamma_{eff}=w_{eff}-1$. In defining the phase space variables we have considered $x=\Omega_{DE}$ and $y=\Omega_{m}$ and $z=\Pi/3H^2$, as it is shown in Eq. \eqref{PhaseSpaceV}. Also, we have defined $\xi_{0}^{*}(\gamma_{DE})=\frac{2\left( {{\gamma}_{DE}}-1\right)}{\sqrt{3}\left( {{{\gamma}_{DE}}^{2}}-2 {{\gamma}_{DE}}-1\right)}$. 
} 
\label{Table1}
\end{table*}

\subsection{Coupling $Q=0$}\label{DyncsA}

\subsubsection{Critical Points and Stability}

For the case $Q=0$, the system of equations \eqref{AutoSys} takes the following form
\bea
&&\frac{dx}{dN}=x\left( 3 x{{\gamma}_{{DE}}}-3 {{\gamma}_{DE}}+3 z-y-4 x+4\right),\nonumber \\
&& \frac{dy}{dN}=3 x y {{\gamma}_{DE}}+3\left( y-1\right) z-{{y}^{2}}+\left( 1-4 x\right) y ,\nonumber \\
&& \frac{dz}{dN}=3 z \left(x {{\gamma}_{DE}}+ {{z}}\right)-z \left(\frac{3 z}{2 y}+\frac{\sqrt{3 y}}{\xi_{0}}\right)+\nonumber\\
&& \left( 1-4 x\right) z-y \left(z+3\right).
\label{AutoSyszeroQ}
\eea 
The autonomous system \eqref{AutoSyszeroQ} admits five critical points which are displayed in Table \ref{Table1}, along with the conditions of existence, acceleration and stability of them.
The analysis of the matrix perturbations, and its eigenvalues for each critical point, are found in Appendix \ref{App1}.

Point $R_{1}$ exist for all values. This critical point is a radiation-dominated solution with $\Omega_{r}=1$, $\Omega_{m}=0$, $\Omega_{DE}=0$ and $\gamma_{eff}=\gamma_{r}=4/3$. By studying the eigenvalues of the matrix of perturbations one finds that it is always a saddle point. 
Points $R_{2\pm}$ exist for all values and both are dark matter-dominated solutions
with $\Omega_{m}=1$, $\Omega_{r}=0$, $\Omega_{DE}=0$. However, due to the viscous pressure of the dark matter fluid the effective EoS parameter is given by
\be
w_{eff}=\gamma_{eff}-1=\frac{1\pm\sqrt{6 {{{\xi}_{0}}^{2}}+1}}{\sqrt{3} {{\xi}_{0}}}.
\ee  which is different from that for the dark matter-dominated era with $\gamma_{m}=1$.
Thus, point $R_{2-}$ presents accelerated expansion for $\xi_{0}>2\sqrt{3}/17$. In the $\xi_{0}\rightarrow 0$ limit one has $w_{eff}\rightarrow -\sqrt{3} \xi_{0}\rightarrow 0^{-}$, and then this solution tends to the dark matter-dominated era with $\gamma_{eff}=\gamma_{m}=1$. From the eigenvalues of the matrix of perturbations, we find that this point is a stable node for $\xi_{0}>\xi_{0}^{*}(\gamma_{DE})$ with  $1-\sqrt{2}< \gamma_{DE}<1$. In another case, for  $\xi_{0}<\xi_{0}^{*}(\gamma_{DE})$ with  $1-\sqrt{2}<\gamma_{DE}<1$, or for $\gamma_{DE}<1-\sqrt{2}$, this point is a saddle point. Given that for point $R_{2+}$ one has $\gamma_{eff}>2/3$, then this point presents only a regime of decelerated expansion for all values of $\xi_{0}$. Furthermore, in this case the effective EoS parameter $w_{eff}$ takes values in the range from $\sqrt{2}$ to $+\infty$, which excludes the dark matter-dominated epoch with $\gamma_{eff}=\gamma_{m}=1$. By studying the corresponding eigenvalues of the matrix of perturbations, we find that this point is always an unstable node. Point $R_{3}$ is a scaling solution with $\Omega_{r}=0$, $\Omega_{m}=\frac{z_{c}}{{{\gamma}_{DE}}-1}$ and $\Omega_{DE}=1-\frac{z_{c}}{{{\gamma}_{DE}}-1}$. It exists for $\gamma_{DE}-1<z_{c}<0$ with $1-\sqrt{2}<\gamma_{DE}<1$. The effective EoS parameter is $\gamma_{eff}=\gamma_{DE}$, and therefore it presents accelerated expansion for $\gamma_{DE}<2/3$. Through the stability analysis we see that this point is a stable node for  $\xi_{0}<\xi_{0}^{*}(\gamma_{DE})$ with  $1-\sqrt{2}<\gamma_{DE}<1$, and a saddle point for  $\xi_{0}>\xi_{0}^{*}(\gamma_{DE})$  with $1-\sqrt{2}<\gamma_{DE}<1$. Finally, Point $R_{4}$ is a dark energy-dominated solution with $\Omega_{DE}=1$ and $\gamma_{eff}=\gamma_{DE}$. Therefore, this point presents accelerated expansion for $\gamma_{DE}<2/3$. From the stability analysis we find that it is a stable node for $-1<\gamma_{DE}<1-\sqrt{2}$ and a stable spiral for $\gamma_{DE}<-1$. On the other hand, for $1-\sqrt{2}<\gamma_{DE}<1$ it is a saddle point. Here, the function $\xi_{0}^{*}(\gamma_{DE})$ is defined in Eq. \eqref{FunczeroQ}.

\begin{figure}[htbp]
\begin{center}
\includegraphics[width=0.5\textwidth,trim=4 4 4 4,clip]{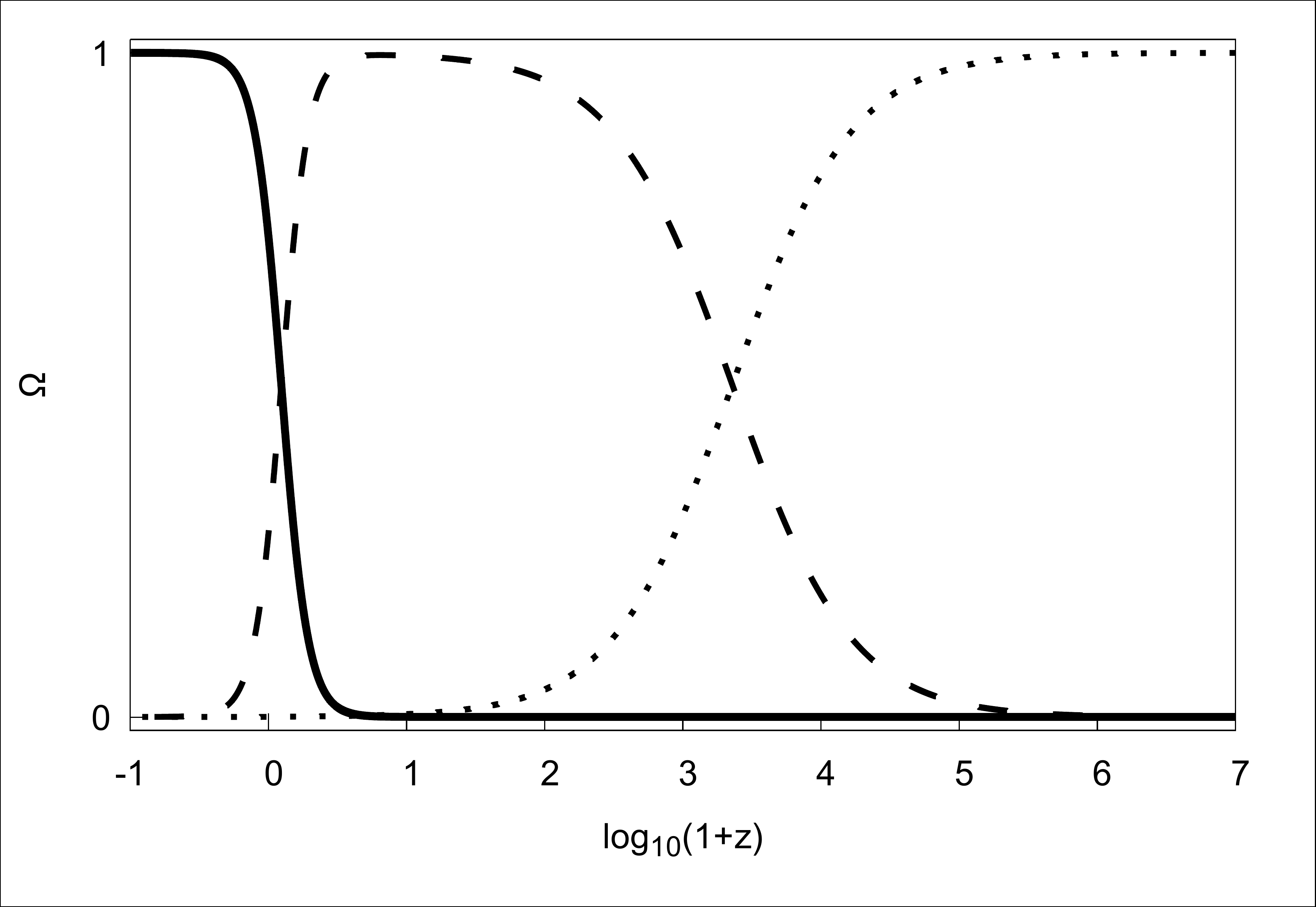}\\
  \includegraphics[width=0.5\textwidth,trim=4 4 4 4,clip]{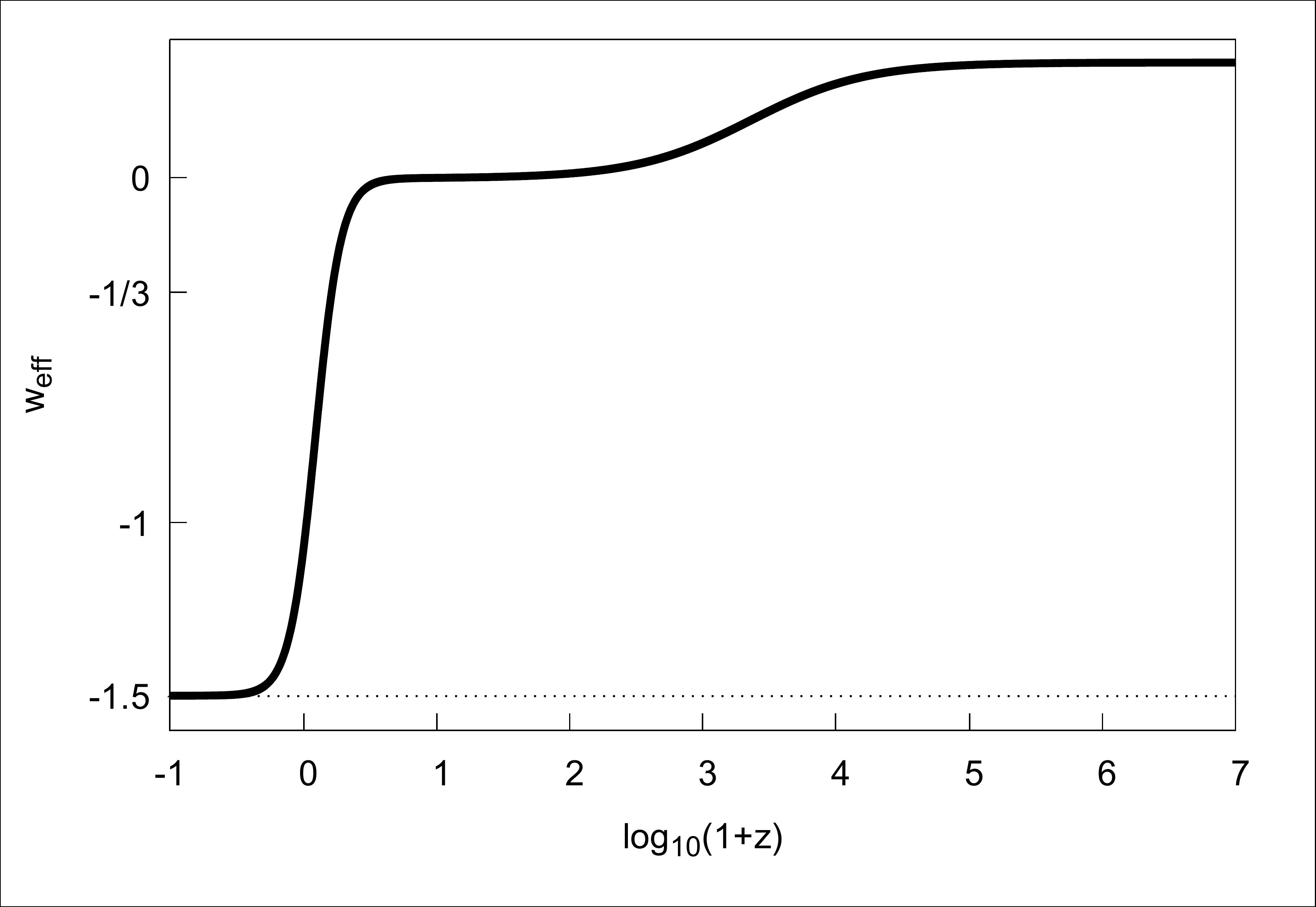}
\end{center}
\caption{\it{Evolution of various observables as a function of the redshift 
($z=\frac{a_0}{a}-1$ and for implicitly we set $a_0=1$) for the dynamical system  \eqref{AutoSyszeroQ} with the values $\xi_{0}=0.001$ and $\gamma_{DE}=-0.5$, in units where $8 \pi G=1$. In the upper graph we depict the evolution of the 
various density parameters, namely $\Omega_{DE}$ (solid line),  $\Omega_{m}$ (dashed 
line), and $\Omega_{r}$ (dotted line). In the lower graph we present the evolution of the effective EoS parameter. The Universe is attracted by the dark energy dominated solution $R_{4}$. For the numerical values we have 
imposed $\Omega_{DE0}\approx 0.72$ and $\Omega_{m0}\approx 0.28$ at present ($z=0$).
Then the effective EoS parameter takes the value $w_{eff}\approx -1.08$ at present, in agreement with observations.}}
\label{FIG1}
\end{figure}

\begin{figure}[htbp]
\begin{center}
\includegraphics[width=0.5\textwidth,trim=4 4 4 4,clip]{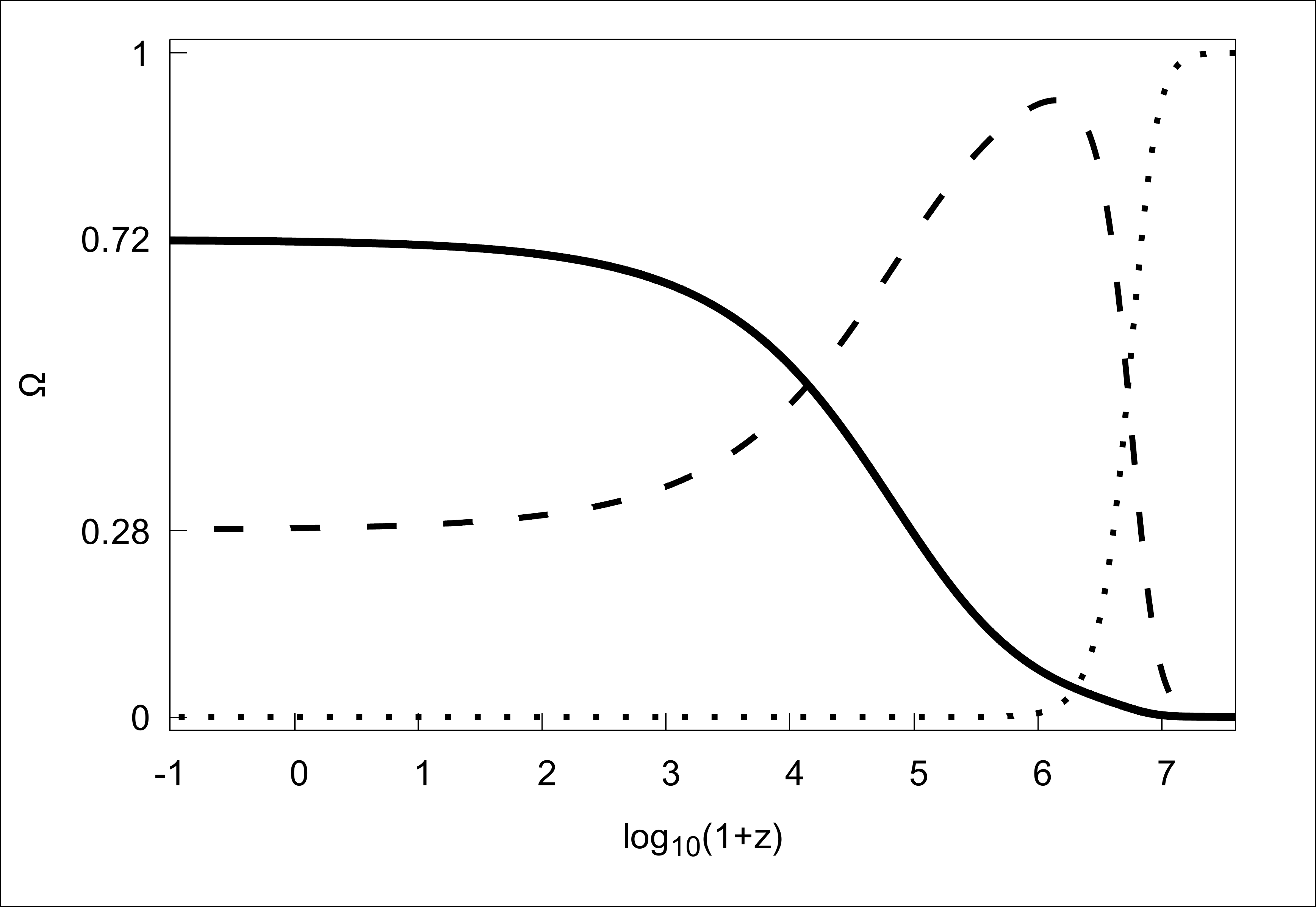}\\
  \includegraphics[width=0.5\textwidth,trim=4 4 4 4,clip]{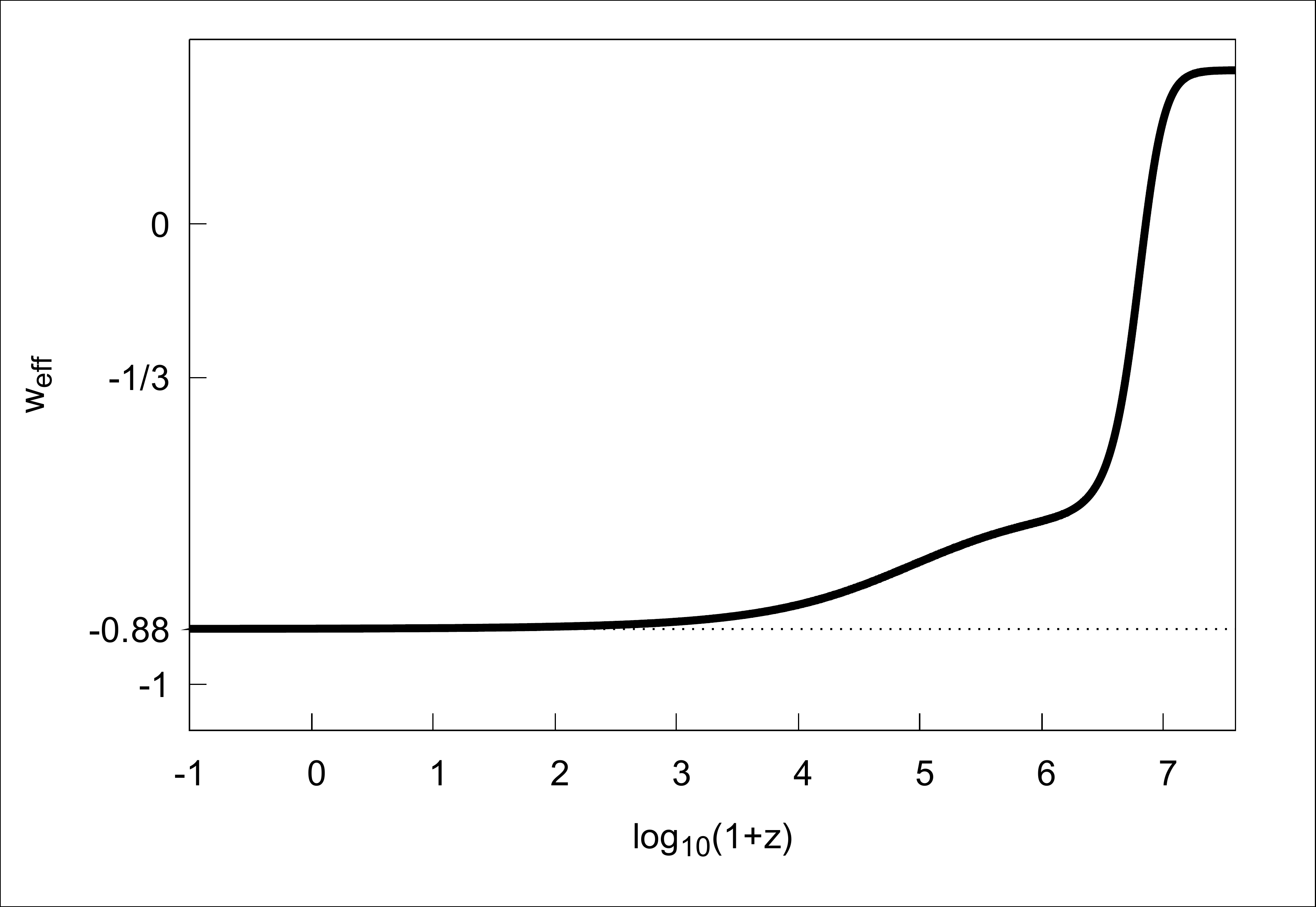}
\end{center}
\caption{\it{Evolution of various observables as a function of the redshift 
($z=\frac{a_0}{a}-1$ and for implicitly we set $a_0=1$) for the dynamical system  \eqref{AutoSyszeroQ} with the values $\xi_{0}=0.44$ and $\gamma_{DE}=0.12$, in units where $8 \pi G=1$. In the upper graph we depict the evolution of the 
various density parameters, namely $\Omega_{DE}$ (solid line),  $\Omega_{m}$ (dashed 
line), and $\Omega_{r}$ (dotted line). In the lower graph we present the evolution of the effective EoS parameter. The Universe is attracted by the scaling solution $R_{3}$. For the numerical values we have 
imposed $\Omega_{DE0}\approx 0.72$ and $\Omega_{m0}\approx 0.28$ at present ($z=0$).
Then the effective EoS parameter takes the value $w_{eff}\approx -0.88$ at present, in agreement with observations. However, the dark-matter-dominated era happens too soon.}}
\label{FIG2}
\end{figure}

\subsubsection{Cosmological Evolution from Critical Points}

In order to obtain a model cosmologically viable it is necessary that it can reproduce the several different epochs in the evolution of the Universe. In the dark matter-dominated era the effective EoS parameter is equal or approximately equal to the equation of state of nonrelativistic matter $w_{m}=\gamma_{m}-1=0$. Any small deviation from this EoS parameter can significantly affect the formation of structures in the Universe. So, point $R_{2+}$ can not reproduce the dark matter-dominated epoch since in this case the effective EoS is very different from $\gamma_{m}=1$. However, point $R_{2-}$ could approximately reproduce a dark matter-dominated era provided that the viscous coefficient $\xi_{0}$ is sufficiently small. In this way, a viable cosmological evolution from these critical points could be either the transition $R_{1} \rightarrow R_{2-} \rightarrow R_{4}$, or, the transition $R_{1} \rightarrow  R_{2-} \rightarrow R_{3}$, provided that $\xi_{0}\ll 1$. In the first case, the Universe tends asymptotically to the accelerated dark-energy-dominated solution $R_{4}$, which presents a phantom nature given that one has in this case $-1<\gamma_{DE}<1-\sqrt{2}$. For instance, in FIG \ref{FIG1} we present the cosmic evolution from the dynamical system \eqref{AutoSyszeroQ} for the values $\xi_{0}=0.001$ and $\gamma_{DE}=-0.5$. The attractor is point $R_{4}$ and we have imposed the conditions $\Omega_{DE0}\approx 0.72$ and $\Omega_{m0}\approx 0.28$ at present time ($z=0$). Then the effective EoS parameter takes the value $w_{eff}=-1.08$ at $z=0$, which shows a phantom-divide crossing during the cosmological evolution. In the second case, the Universe tends asymptotically to the scaling solution $R_{3}$ with $0<\Omega_{DE}< 1$, which is also an accelerated solution for $\gamma_{DE}<2/3$. This solution includes the cases of a cosmological constant-like ($\gamma_{DE}=0$) and quintessence-like ($0<\gamma_{DE}<1$) solutions, with $\Omega_{DE}\approx 1$ and $\xi_{0}\lesssim 0.01 <\xi_{0}^{*}(\gamma_{DE})$. Moreover, the most interesting on this solution is that it can provide a way to alleviate the cosmological coincidence problem \cite{AmendolaTsujikawa,Wang:2016lxa,Quartin:2008px}. However, in alleviating the cosmic coincidence problem via this scaling solution, it is necessary to assume large values of the viscosity coefficient $\xi_{0}$ which affects the dark matter-dominated era and therefore the structures formation in the Universe. To see this more clearly from some numerical results, let us introduce the constant $0<C<\sqrt{4/3}$, such that $x_{c}=1-3C^2/4$, $y_{c}=3C^2/4$ and $z_{c}=3 C^2(\gamma_{DE}-1)/4$. In this way, we have that $C$, $\gamma_{DE}$ and $\xi_{0}$ must satisfy the relation $\xi_{0}=C(\gamma_{DE}-1)/(\gamma_{DE}^2-2\gamma_{DE}-1)$. For example, for $C\approx 0.61$ ( $x_{c} \approx 0.72$ and $y_{c}\approx 0.28$) and $\gamma_{DE}<0.2$, the viscous coefficient must satisfy the constraint $\xi_{0}>0.36$, which is disagree with the physical requirement $\xi_{0}\ll 1$. In FIG \ref{FIG2} we shows the better result for the attractor $R_{3}$ such that in the present time the energy fractions are $\Omega_{DE}\approx 0.72$ and $\Omega_{m}\approx 0.28$. In this case, although the effective EoS parameter tends to the value $w_{eff}=\gamma_{eff}-1\approx-0.88$, well within the required range in accordance to observational data, the dark-matter-dominated era happens too soon. As we will see below, the introduction of an explicit coupling between dark energy and viscous dark matter allows to obtain another scaling solution which provides a viable cosmological evolution and at the same time alleviating the cosmological coincidence problem.  

\def\tablename{Table}%
\begin{table*}[ht]
\centering
\begin{center}
\begin{tabular}{c c c c c c c c}\hline\hline
Name &   $x_{c}$& $y_{c}$ & $z_{c}$ &  Existence & $\gamma_{eff}$&  Acceleration & 
Stability 
\\\hline 
\\ 
$S_{1}$ & $0$ & $0$ & $0$ & Always & $\frac{4}{3}$  & Never  & Saddle point for \\&&&&&&&  all values  \\ \\ 
$S_{2-}$ & $0$ & $1$ & $\frac{1-\sqrt{6 {{{\xi}_{0}}^{2}}+1}}{\sqrt{3} {{\xi}_{0}}}$ & Always  &  $1+\frac{1-\sqrt{6 {{{\xi}_{0}}^{2}}+1}}{\sqrt{3} {{\xi}_{0}}}$  & $\xi_{0}>\frac{2\sqrt{3}}{17}$ & Stable node for \\&&&&&&&  $\xi_{0}>\xi_{0}^{*}(\gamma_{DE},\alpha)$  with \\&&&&&&&  $-\sqrt{2}+1<\gamma_{DE}+\alpha<1$ \\&&&&&&& and  $1-\sqrt{2}<\alpha<1+\sqrt{2}$, or\\ &&&&&&&
for all values of $\xi_{0}$ and \\ &&&&&&&
 $1-\alpha<\gamma_{DE}<1$ \\ &&&&&&& 
Saddle point for \\&&&&&&& $\xi_{0}<\xi_{0}^{*}(\gamma_{DE})$ with \\&&&&&&&  $-\sqrt{2}+1<\gamma_{DE}+\alpha<1$ \\&&&&&&&
and $1-\sqrt{2}<\alpha<1+\sqrt{2}$ or \\ &&&&&&&
for all values of $\xi_{0}$ with \\&&&&&&& 
$\gamma_{DE}<-\alpha-\sqrt{2}+1$ \\&&&&&&& 
and $1-\sqrt{2}<\alpha<1+\sqrt{2}$ or \\&&&&&&&
for all values of $\xi_{0}$ with \\&&&&&&&
$\gamma_{DE}<1$ and $\alpha<1-\sqrt{2}$ \\ \\
$S_{2 +}$ 
& $0$ & $1$ & $\frac{1+\sqrt{6{{{\xi}_{0}}^{2}}+1}}{\sqrt{3}{{\xi}_{0}}}$ &  Always & $1+\frac{1+\sqrt{6{{{\xi}_{0}}^{2}}+1}}{\sqrt{3}{{\xi}_{0}}}$& Never & Unstable node for \\&&&&&&&  
$\gamma_{DE}<1-\alpha$ or \\&&&&&&&  
for $1<\gamma_{DE}+\alpha<1+\sqrt{2}$ \\&&&&&&& 
 with $0<\alpha<\sqrt{2}+1$\\ \\
$S_{3}$ 
& $1-\frac{3 {{C}^{2}}}{4}$ & $\frac{3 {{C}^{2}}}{4}$ & $\frac{3 {{C}^{2}}\left( {{\gamma}_{DE}}-1\right)}{4}+\alpha$ & $0<C<\sqrt{4/3}$ \\&&&&   with $\gamma_{DE}<1$ & ${{\gamma}_{DE}}+\alpha$  & $\gamma_{DE}<\frac{2}{3}-\alpha$ &  Stable for\\&&&&&&&  $\mathcal{E}<0$ and $\mathcal{F}>0$ with \\&&&&&&& $\gamma_{DE}<4/3-\alpha$
\\&&&&&&& Unstable for \\&&&&&&& $\mathcal{E}>0$ or  $\mathcal{F}<0$ or \\&&&&&&& $\gamma_{DE}>4/3-\alpha$
\\\\
\hline\hline
\end{tabular}
\end{center}
\caption{The physical critical points of the dynamical system  \eqref{AutoSysnonzeroQ} with coupling $Q=3\alpha H \rho_{DE}$, and their existence and stability conditions, along with the  conditions for 
accelerated expansion, and the corresponding values of the effective barotropic parameter $\gamma_{eff}=w_{eff}-1$.  In defining the phase space variables we have considered $x=\Omega_{DE}$ and $y=\Omega_{m}$ and $z=\Pi/3H^2$, as it is shown in Eq. \eqref{PhaseSpaceV}. For point $S_{-}$ we have defined $\xi_{0}^{*}(\gamma_{DE},\alpha)=\frac{2 \left( {{\gamma}_{DE}}+\alpha-1\right) }{\sqrt{3}\left( {{{\gamma}_{DE}}^{2}}+2\alpha {{\gamma}_{DE}}-2 {{\gamma}_{DE}}+{{\alpha}^{2}}-2\alpha-1\right)}$. On the other hand, for point $S_{3}$ we have defined $\mathcal{E}=3 C^2\left[ {{\xi}_{0}}\left(2 {{\gamma}_{DE}}+ 3 \alpha-2\right)- {{C}}\right]-4{{\xi}_{0}}\alpha$ and $\mathcal{F}=3 C\gamma_{DE}\left[2 {{\xi}_{0}}\left( {{{\gamma}_{DE}}}+ \alpha-2\right)-3 {{C}}\right]+ 9 {{C}^{2}}-6 \xi_{0}\left(\alpha+1\right) C-4\alpha$, with the parameters, $\xi_{0}$, $C$, $\gamma_{DE}$ and $\alpha$, satisfying the relation \eqref{Alpha}.} 
\label{Table2}
\end{table*}

\begin{figure}[htbp]
\begin{center}
\includegraphics[width=0.5\textwidth,trim=4 4 4 4,clip]{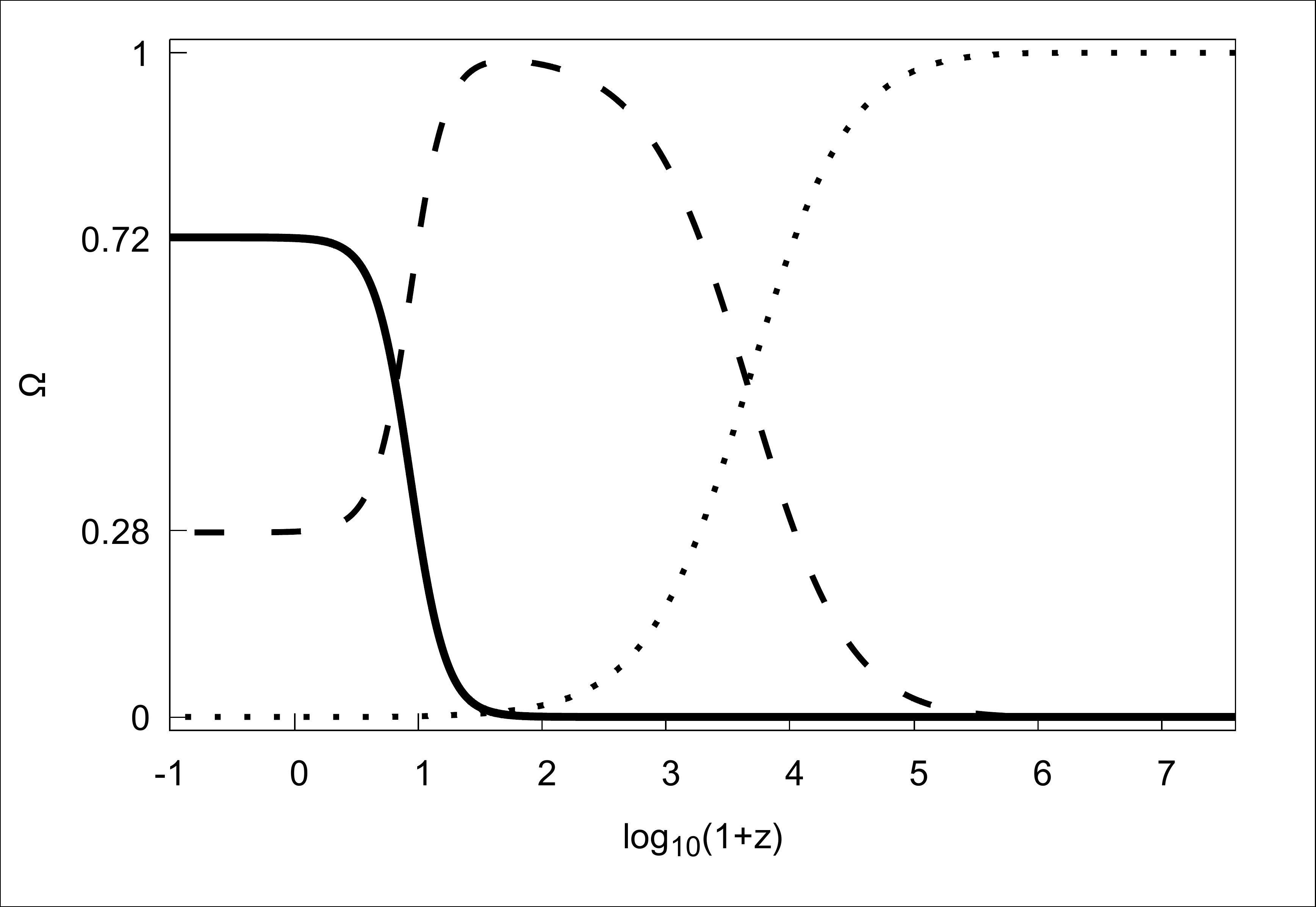}\\
  \includegraphics[width=0.5\textwidth,trim=4 4 4 4,clip]{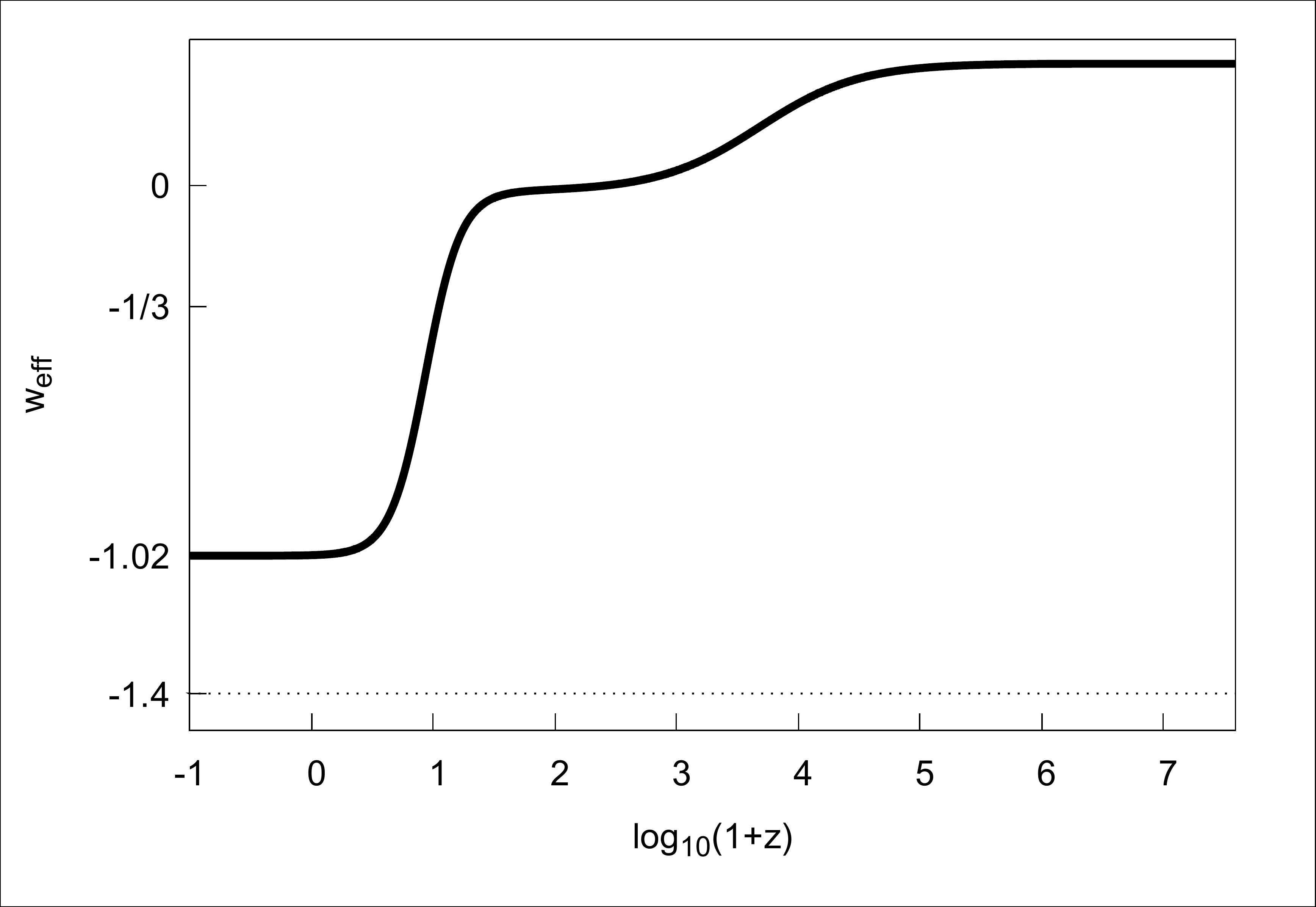}
\end{center}
\caption{\it{Evolution of various observables as a function of the redshift 
($z=\frac{a_0}{a}-1$ and for implicitly we set $a_0=1$) for the dynamical system  \eqref{AutoSysnonzeroQ} with the values $\xi_{0}=0.01$, $\gamma_{DE}=-0.4$, $C=0.61$ and $\alpha=0.38$ in units where $8 \pi G=1$. In the upper graph we depict the evolution of the 
various density parameters, namely $\Omega_{DE}$ (solid line),  $\Omega_{m}$ (dashed 
line), and $\Omega_{r}$ (dotted line). In the lower graph we present the evolution of the effective EoS parameter. The Universe is attracted by the scaling solution $S_{3}$. For the numerical values we have 
imposed $\Omega_{DE0}\approx 0.72$ and $\Omega_{m0}\approx 0.28$ at present ($z=0$).
Then the effective EoS parameter takes the value $w_{eff}\approx -1.02$ at present, in agreement with observations. }}
\label{FIG3}
\end{figure}

\subsection{Coupling $Q=3\alpha H \rho_{DE}$}\label{DyncsB}
\subsubsection{Critical points and Stability}

In the case of a nonzero coupling in the form $Q=3\alpha H \rho_{DE}$, we have the following autonomous system 
\bea
&&\frac{dx}{dN}=x\left( 3 x{{\gamma}_{{DE}}}-3 {{\gamma}_{DE}}+3 z-y-4 x+4-3 \alpha\right),\nonumber \\
&& \frac{dy}{dN}=3 x y {{\gamma}_{DE}}+3\left(y-1\right) z-{{y}^{2}}+\left( 1-4 x\right) y +3 \alpha x ,\nonumber \\
&& \frac{dz}{dN}=3 z \left(x {{\gamma}_{DE}}+ {{z}}\right)-z \left(\frac{3 z}{2 y}+\frac{\sqrt{3 y}}{\xi_{0}}\right)+\nonumber\\
&& \left( 1-4 x\right) z-y \left(z+3\right)+\frac{3\alpha x z}{2 y}.
\label{AutoSysnonzeroQ}
\eea 
All the critical points of the dynamical system \eqref{AutoSysnonzeroQ}, along with the existence, accelerated and stability conditions, are displayed in Table \ref{Table2}. Also, the analysis of the matrix perturbations, and its eigenvalues for each critical point, are found in Appendix \ref{App2}.

For $x=0$ we obtain the same critical points as for the case $Q=0$. However, the stability conditions for them are altered due to the presence of a nonzero coupling parameter $\alpha$. Point $S_{1}$ is a radiation-dominated solution with $\Omega_{r}=1$, $\Omega_{m}=0$, $\Omega_{DE}=0$ and effective EoS parameter $\gamma_{eff}=\gamma_{r}=4/3$. This critical point is a saddle point for all the values. Points $S_{2\pm}$ are dark matter-dominated solutions with $\Omega_{r}=0$, $\Omega_{m}=1$ and $\Omega_{DE}=0$. However, for these points one find that the effective EoS parameter depends on the viscosity coefficient $\xi_{0}$, and therefore they are different from the pressureless matter with $\gamma_{m}=1$. Point $S_{2+}$ is always an unstable node, whereas point $S_{2-}$ is a saddle point and also a stable node. On the other hand, from Eqs. \eqref{AutoSysnonzeroQ} and for $x\neq 0$ we find the following family of critical points 
\be
x_{c}=1-\frac{z_{c}-\alpha}{{{\gamma}_{\mathit{DE}}}-1},\:\:\:\: y_{c}=\frac{z_{c}-\alpha}{{{\gamma}_{\mathit{DE}}}-1}, 
\ee and $z_{c}$ satisfying the relation
\be
\sqrt{\frac{z_{c}-\alpha}{3\left({{\gamma}_{DE}}-1\right)}}=\frac{\xi_{0}}{2}\left[{{\gamma}_{DE}}+\alpha-1-\frac{2\left(z_{c}-\alpha\right)}{z_{c}\left( {{\gamma}_{DE}}-1\right) }\right],
\label{Constr1}
\ee along with the physical constraint ${{\gamma}_{\mathit{DE}}}+\alpha-1<z_{c}<\alpha$ for $\gamma_{DE}<1$. This system of equations is more complicated than for the case $Q=0$ and it does not easy to solve for $z_{c}$ from Eq.~\eqref{Constr1}. Nevertheless, we are interested in the scaling solution with $x_{c}=1-3 C^2/4$, $y_{c}=3 C^2/4$ and $z_{c}=3 C^2\left(\gamma_{DE}-1\right)/4+\alpha$, such that $0<C<\sqrt{4/3}$. For this scaling solution we find that the effective EoS parameter is given by  $\gamma_{eff}=\gamma_{DE}+\alpha$, and therefore it presents accelerated expansion for $\gamma_{DE}<2/3-\alpha$.
On the other hand, from \eqref{Constr1} we obtain that the parameters $\xi_{0}$, $\gamma_{DE}$, $\alpha$ and $C$, must satisfy the relation
\be
\alpha=\frac{3 {{C}^{2}} \left[C\left( {{\gamma}_{eff}}-1\right)-{{\xi}_{0}}\left({{{\gamma}_{eff}}^{2}}-2{{\gamma}_{eff}}-1\right)\right]}{\left[C+\xi_{0}\left(1-{{\gamma}_{eff}}\right)\right]\left( 3 {{C}^{2}}-4\right) },
\label{Alpha}
\ee where $\gamma_{eff}=\gamma_{DE}+\alpha$. So, given the values of $\gamma_{eff}$, $\xi_{0}$ and $C$, we can obtain from Eq. \eqref{Alpha} the required values of $\alpha$ and $\gamma_{DE}$, such that the solution is a fixed point of the autonomous system \eqref{AutoSysnonzeroQ}. From the corresponding stability analysis we find that this critical point can be an attractor in the required ranges for the parameters (see Appendix \ref{App2}).

\subsubsection{Cosmological Evolution from Critical points}

Since the critical point $S_{2-}$ can approximately reproduce a dark matter-dominated era provided that the viscous coefficient $\xi_{0}$ is sufficiently small, then a viable cosmological evolution from the above critical points is determined by the transition $S_{1} \rightarrow S_{2-} \rightarrow S_{3}$. In this case the final attractor is the scaling solution $S_{3}$, which presents accelerated expansion and also can alleviate the cosmological coincidence problem. This  solution $S_{3}$ generalizes the scaling solution $R_{3}$ for the case of a nonzero coupling $\alpha$ between dark energy and viscous dark matter. This coupling allows to obtain from point $S_{3}$ a viable cosmological evolution, without spoiling the large-scale structures formation of the Universe, and at the same time alleviating the cosmological coincidence problem. Furthermore, in this case a phantom nature of dark energy is favored. For example, for $C=0.61$ ($x_{c}\approx 0.72$, $y_{c}\approx 0.28$), $\xi_{0}=0.01$, and an effective EoS parameter $w_{eff}$ in the physical range $(-1.38,-0.89)$, in accordance with the latest Planck results \cite{Ade:2013zuv}, we obtain that the constant coupling $\alpha$ must be in the range $(0.33,0.52)$ and the dark energy EoS parameter $\omega_{DE}$ must be in the phantom range $(-1.9,-1.22)$.
In FIG \ref{FIG3} we show the cosmological evolution when the final attractor is point $S_{3}$, and fix the energy fractions to be $\Omega_{DE}\approx 0.72$ and $\Omega_{m}\approx 0.28$ in the present time $z=0$. The thermal history of the Universe is successfully reproduced, and the effective EoS parameter tends to value $w_{eff}\approx -1.02$ well within the requested range by observations.

%%%%%%%%%%%%%%%%%%%%%%%%%%%%%%%%%%%%%%%%%%%%%%%%%%%%%%%%%%%%
\section{Concluding Remarks}\label{Conclusions}

From the viewpoint of the fluid description, and the current cosmological observational data, there is not reason for excluding a more general setting where the energy components of the Universe are treated as imperfect fluids due to the presence of a bulk viscosity in them. This bulk viscosity is introduced in the energy conservation equation through an additional term $\Pi$, which represents the viscous pressure of the fluid. The nature of this viscous pressure $\Pi$ is determined by the class of formalism or approach adopted in the description of the viscous fluid. There are two classes of formalisms which are generally used in the description of viscous fluids in Cosmology.
The first one is the noncausal Eckart formalism \cite{Eckart:1940zz}, which has been 
applied by many authors in the study of viscous cosmologies in a flat FRW Universe, in both the background \cite{Velten:2013qna, Leyva:2016gzm, Cruz:2014iva, Avelino:2013wea}, as well as in the perturbation level \cite{Velten:2014xca,Velten:2013pra,Velten:2013rra,Velten:2012uv}.  The second one is the causal Israel-Stewart (IS) formalism \cite{Israel:1979wp}, in which is introduced a transport equation for the viscous pressure, allowing a noninstantaneous propagation of the interactions, and in this way eliminating the drawbacks related to causality and stability of the Eckart formalism  \cite{Maartens:1996vi,Cruz:2016rqi}. 

In the present paper we have performed a detailed dynamical analysis for viscous cosmologies in the framework of the causal IS formalism. In particular, we deepen the study of ``Scaling Solutions''. Dynamical systems in Cosmology have a very important role in the study of the cosmological viability of a given model in the background level \cite{Coley:2003mj}. Moreover, scaling solutions are attractors which provide a new line of attack on the fine-tuning problem, or cosmological coincidence problem of dark energy \cite{Copeland:2006wr,AmendolaTsujikawa}.
In our study we have considered the total cosmic fluid constituted by radiation, dark matter and dark energy. The dark matter fluid is treated as an imperfect fluid which has a bulk viscosity that depends on its energy density in the usual form $\xi(\rho_{m})=\xi_{0} \rho_{m}^{1/2}$, whereas the other components are assumed to behave as perfect fluids with constant EoS parameter. Here, we also have considered the possibility of an explicit coupling $Q$ between dark energy and viscous dark matter. Following the literature, this coupling is assumed to be proportional to the dark energy density $\rho_{DE}$, and the proportionality coefficient being normalized by the Hubble expansion rate $H$, in the form $Q=3\alpha H \rho_{DE}$ with $\alpha$ the coupling constant \cite{Copeland:2006wr}.

We have shown that the thermal history of the Universe is reproduced provided that the viscous coefficient satisfies the condition $\xi_{0}\ll 1$, either for a zero or a suitable nonzero coupling. In this case, the final attractor is a dark-energy-dominated, accelerating Universe, with effective EoS parameter in the quintessence-like, cosmological constant-like or phantom-like regime, in agreement with observations. As our main result, we have shown that in order to obtain a viable cosmological evolution and at the same time alleviating the cosmological coincidence problem via the mechanism of scaling solution, an explicit interaction between dark energy and viscous dark matter seems inevitable. This result is consistent with the well-known fact that models where dark matter and dark energy interact with each other have been proposed to solve the coincidence problem \cite{Wang:2016lxa,Quartin:2008px}. However, it is important note that several other relevant efforts have been performed in showing that a combination of a dark matter fluid with a bulk dissipative pressure and a quintessence field can simultaneously drive an accelerated expansion phase and solve the cosmological coincidence problem without the need for an explicit coupling between dark energy and viscous dark matter \cite{Chimento:2000kq,Chimento:2002ej,Ibanez:2002ra,Ibanez:2003hj}.

In the case of a zero coupling ($Q=0$) between dark energy and viscous dark matter, the scaling attractor is the critical point $R_{3}$ in Table \ref{Table1}. For $x_{c}=\Omega_{DE} \approx 0.72$, $y_{c}=\Omega_{m}\approx 0.28$ and $\gamma_{DE}<0.2$, the viscous coefficient must satisfy the constraint $\xi_{0}>0.36$, which is disagree with the physical requirement $\xi_{0}\ll 1$. On the other hand, by considering a nonzero coupling in the form $Q=3\alpha H \rho_{DE}$, we find that the scaling attractor is the critical point $S_{3}$ in Table \ref{Table2}, which allows a viable cosmological evolution and at the same time alleviating the cosmological coincidence problem.
For example, from this scaling solution and Eq. \eqref{Alpha}, for $x_{c}=\Omega_{DE}\approx 0.72$, $y_{c}=\Omega_{m}\approx 0.28$, $\xi_{0}= 0.01$, and an effective EoS parameter $w_{eff}$ in the physical range $(-1.38,-0.89)$, in accordance with the latest Planck results \cite{Ade:2013zuv}, we obtain that the constant coupling $\alpha$ lies in the tight range $(0.33,0.52)$, whereas the dark energy EoS parameter $w_{DE}$ must be in the phantom range $(-1.9,-1.22)$. Thus, in the case of a nonzero coupling between dark energy and dark matter, $w_{eff}$ and $w_{DE}$ take different values due to this coupling.
In our model, $w_{eff}$ must be in accordance with observational data, whereas $w_{DE}$ should be determined from the values for $w_{eff}$ by using Eq. \eqref{Alpha} and the relation $w_{eff}=w_{DE}+\alpha$. In FIG \ref{FIG3}, we have presented the behavior of the cosmological parameters when the final attractor is the scaling solution $S_{3}$, for specific values of the free parameters in the requested ranges. 

Here it is worth noting that the critical points $\{R_{1},R_{2\pm},R_{3},R_{4}\}$ displayed in Table \ref{Table1}, constitute a causal extension of the corresponding critical points $\{P_{1},P_{2},P_{3},P_{4}\}$ obtained in the context of the noncausal Eckart formalism and displayed in Table I of Ref. \cite{Cruz:2014iva}. In the limit of a very small relaxation time $\tau$, or equivalently for $\xi_{0} \rightarrow 0$, we recover the matter solution $P_{2}$ from critical point $R_{2-}$. The numerical difference between the values of these two matter solutions is determined by the values of $\xi_{0}$, and it should be constrained from observational data. As expected, for late-times the difference between these set of fixed points is even more pronounced. From the scaling attractor $R_{3}$ does not possible to obtain the scaling attractor $P_{3}$ only by taking the $\xi_{0} \rightarrow 0$ limit, since in this case $R_{3}$ depends also on the  dark energy EoS parameter. A similar situation occurs for the dark energy-dominated solution $R_{4}$, which is an attractor in a more wide range of parameters, whereas the corresponding point $P_{4}$ is an attractor only for $\xi_{0}=0$.  Moreover, we can also mention that the critical points $\{S_{1},S_{2\pm},S_{3}\}$ in Table \ref{Table2} generalize the above critical points for the case of a suitable nonzero coupling between dark energy and viscous dark matter.

Finally, we mention that in studying the dynamics of any model we are interest in the cosmological properties and stability conditions of late-time asymptotic solutions. This study is important because it provides us an initial test for the viability of a given model in reproducing the thermal history of the Universe. However, in order to decide whether a theory is or not a candidate for the description of Nature, many relevant investigations are necessary. Apart from the correct late-time, asymptotic, behaviour, it must be able to describe the correct universe evolution at early and intermediate times too. In particular, one should perform a detailed comparison with observational data, as for instance SNIa, BAO, CMB, and large-scale structure (LSS) \cite{AmendolaTsujikawa}.

%%%%%%%%%%%%%%%%%%%%%%%%%%%%%%%%%%%%%%%%%%%%%%%%%%%%%%%%%%%%

%%%%%%%%%%%%%%%%%%%%%%%%%%%%%%%%%%%%%%%%%%%%%%%%%%%%%%%%%%%%%%%%%%%%
\begin{acknowledgments}
G. Otalora acknowldeges DI-VRIEA for financial support through Proyecto Postdoctorado $2017$ VRIEA-PUCV.
\end{acknowledgments}
%%%%%%%%%%%%%%%%%%%%%%%%%%%%%%%%%%%%%%%%%%%%%%%%%%%%%%%%%%%%%%%%%%%%

\appendix

\section{Stability Analysis}

\subsection{Case $Q=0$}\label{App1}

In order to examine the stability of the critical points, we perform linear 
perturbations around them as $x_{i}=x_{i}^{*}+\delta{x_{i}}$ and thus we extract the perturbation equations as  $ \textbf{U}'={\mathcal{M}}\cdot \textbf{U}$, where $\textbf{U}$ is the
column vector of the perturbations $\delta{x_{i}}$ and $\mathcal{M}$ is the  $3\times3$  matrix that contains the coefficients of the perturbation equations. For the autonomous system \eqref{AutoSyszeroQ} the nonzero components of the matrix of perturbations are
\bea
&& M_{11}=3\left( 2 x-1\right){{\gamma}_{DE}}+3 z-y-8 x+4,\\
&& M_{12}=-x,\\
&& M_{13}=3 x,\\
&& M_{21}=y\left(3{{\gamma}_{DE}}-4\right),\\ 
&& M_{22}=3 x {{\gamma}_{DE}}+3 z-2 y-4 x+1,\\
&& M_{23}=3\left(y-1\right),\\
&& M_{31}=z \left( 3 {\gamma}_{DE}-4\right),\\
&& M_{32}=\frac{\sqrt{3} z \left( \sqrt{3} {{\xi}_{0}} z-{{y}^{\frac{3}{2}}}\right) }{2 {{\xi}_{0}}{{y}^{2}}}-z-3,\\
&& M_{33}=x\left( 3 {{\gamma}_{DE}}-4\right) -\frac{\sqrt{3}\left( \sqrt{3}{{\xi}_{0}}z+{{y}^{\frac{3}{2}}}\right) }{{{\xi}_{0}} y}+\nonumber\\
&& 6 z-y+1.
\eea

Concerning critical point $R_{1}$, the corresponding eigenvalues are
given by 
\bea
&&\mu_{1}=\frac{1-\sqrt{35}}{2},\\
&& \mu_{2}=\frac{\sqrt{35}+1}{2}, \\
&& \mu_{3}=4-3 {{\gamma}_{DE}}.\\
\eea Critical point $R_{1}$ is a saddle point.

For the critical points $R_{2-}$ and $R_{2+}$ the eigenvalues are 
\bea
&& {{\mu}_{1}}=3\left(1-{{\gamma}_{DE}}\right)+\frac{\sqrt{3}\left(1\pm \sqrt{6 {{{\xi}_{0}}^{2}}+1}\right)}{{{\xi}_{0}}},\\
&& \mu_{2}=-1+\frac{\sqrt{3}\left(1\pm \sqrt{6 {{{\xi}_{0}}^{2}}+1}\right)}{{{\xi}_{0}}},\\
&& \mu_{3}=\pm\frac{\sqrt{3}\sqrt{6{{{\xi}_{0}}^{2}}+1}}{{{\xi}_{0}}}.
\eea Critical point $R_{2-}$ is a stable node for $\xi_{0}>\xi_{0}^{*}(\gamma_{DE})$ with  $1-\sqrt{2}<\gamma_{DE}<1$. It is a saddle point for $\xi_{0}<\xi_{0}^{*}(\gamma_{DE})$ with  $1-\sqrt{2}<\gamma_{DE}<1$, or for $\gamma_{DE}<1-\sqrt{2}$.  Here we have defined the function
\be
\xi_{0}^{*}(\gamma_{DE})=\frac{2\left( {{\gamma}_{DE}}-1\right)}{\sqrt{3}\left( {{{\gamma}_{DE}}^{2}}-2 {{\gamma}_{DE}}-1\right)}, 
\label{FunczeroQ}
\ee with $1-\sqrt{2}<\gamma_{DE}<1$. On the other hand, critical point $R_{2+}$ is an unstable node for all values. 

Critical point $R_{3}$ has eigenvalues given by
\bea
&&
{{\mu}_{1}}=\frac{3\left( {{\gamma}_{DE}}-1\right)\left(\frac{2}{{{\left( {{\gamma}_{DE}}-1\right) }^{2}}}+1\right)}{4}\times \nonumber\\
&& \Bigg[1-\sqrt{1-\frac{\mathcal{A}\left( {{{\gamma}_{DE}}^{2}}-2{{\gamma}_{DE}}-1\right) }{{{\left( {{\gamma}_{DE}}-1\right) }^{4}} {{\left( \frac{2}{{{\left( {{\gamma}_{DE}}-1\right) }^{2}}}+1\right) }^{2}}}}\Bigg],\\
&&{{\mu}_{2}}=\frac{3\left( {{\gamma}_{DE}}-1\right)\left(\frac{2}{{{\left( {{\gamma}_{DE}}-1\right) }^{2}}}+1\right)}{4}\times \nonumber\\
&& \Bigg[1+\sqrt{1-\frac{\mathcal{A}\left( {{{\gamma}_{DE}}^{2}}-2{{\gamma}_{DE}}-1\right) }{{{\left( {{\gamma}_{DE}}-1\right) }^{4}} {{\left( \frac{2}{{{\left( {{\gamma}_{DE}}-1\right) }^{2}}}+1\right) }^{2}}}}\Bigg],\\
&&\mu_{3}=3 {{\gamma}_{DE}}-4, 
\eea where we have defined the quantity
\bea
&& \mathcal{A}=\gamma_{DE}^2\left[3\xi_{0}^2\gamma_{DE}\left({{{\gamma}_{DE}}}-4\right)+2\left(3 {{{\xi}_{0}}^{2}}-2\right)\right]+\nonumber\\
&& 4 \left( 3 {{{\xi}_{0}}^{2}}+2\right) {{\gamma}_{DE}}+3 {{{\xi}_{0}}^{2}}-4. 
\eea In the range of definition this critical point is a stable node for  $\xi_{0}<\xi_{0}^{*}(\gamma_{DE})$ and  $1-\sqrt{2}<\gamma_{DE}<1$. Also, it is a saddle point for $\xi_{0}>\xi_{0}^{*}(\gamma_{DE})$ and $1-\sqrt{2}<\gamma_{DE}<1$.

Finally, for critical point $R_{4}$ we find the eigenvalues
\bea
&&{{\mu}_{1}}=\frac{3\left(\gamma_{DE}-1\right)}{2}\times\Bigg[1+\nonumber\\
&& \sqrt{1-2\frac{\left({{{\gamma}_{DE}}^{2}}-2{{\gamma}_{DE}}-1\right)}{\left({{\gamma}_{DE}}-1\right)^2 }}\Bigg],\\
&& {{\mu}_{2}}=\frac{3\left(\gamma_{DE}-1\right)\left(1-\sqrt{1-2\frac{\left({{{\gamma}_{DE}}^{2}}-2{{\gamma}_{DE}}-1\right)}{\left({{\gamma}_{DE}}-1\right)^2 }}\right)}{2},\\
&&\mu_{3}=3 {{\gamma}_{DE}}-4.
\eea  This critical point is a stable node for  $-1<\gamma_{DE}<1-\sqrt{2}$. For $\gamma_{DE}<-1$ it is a stable spiral. Saddle point for $1-\sqrt{2}<\gamma_{DE}<1$.

\subsection{Case $Q=3 \alpha H \rho_{DE}$, $\abs{\alpha}\leq 1$}
\label{App2}

Perturbing to linear order around each critical point the dynamical system \eqref{AutoSysnonzeroQ}, we obtain the matrix of perturbations  with the following nonzero components
\bea
&& M_{11}=3\left( 2 x-1\right){{\gamma}_{DE}}+3 z-y-8 x+4-3 \alpha,\\
&& M_{12}=-x,\\
&& M_{13}=3 x,\\
&& M_{21}=y\left(3{{\gamma}_{DE}}-4\right)+3 \alpha,\\ 
&& M_{22}=3 x {{\gamma}_{DE}}+3 z-2 y-4 x+1,\\
&& M_{23}=3\left(y-1\right),\\
&& M_{31}=z \left( 3 {\gamma}_{DE}-4\right)+\frac{3 \alpha z}{2 y},\\
&& M_{32}=\frac{\sqrt{3} z \left( \sqrt{3} {{\xi}_{0}} z-{{y}^{\frac{3}{2}}}\right) }{2 {{\xi}_{0}}{{y}^{2}}}-z-3-\frac{3 \alpha x z}{2 {{y}^{2}}},\\
&& M_{33}=x\left( 3 {{\gamma}_{DE}}-4\right) -\frac{\sqrt{3}\left( \sqrt{3}{{\xi}_{0}}z+{{y}^{\frac{3}{2}}}\right) }{{{\xi}_{0}} y}+\nonumber\\
&& 6 z-y+1 +\frac{3\alpha x}{2 y}.
\eea

For critical point $S_{1}$ we find the eigenvalues
\bea
&&\mu_{1}=\frac{1-\sqrt{35}}{2},\\
&& \mu_{2}=\frac{\sqrt{35}+1}{2}, \\
&& \mu_{3}=4-3\left({{\gamma}_{DE}}+\alpha\right).
\eea  Then this critical point is a saddle point for all values. 

For the critical points $S_{2-}$ and $S_{2+}$ the eigenvalues are 
\bea
&& {{\mu}_{1}}=3\left(1-{{\gamma}_{DE}}-\alpha\right)+\frac{\sqrt{3}\left(1\pm \sqrt{6 {{{\xi}_{0}}^{2}}+1}\right)}{{{\xi}_{0}}},\\
&& \mu_{2}=-1+\frac{\sqrt{3}\left(1\pm \sqrt{6 {{{\xi}_{0}}^{2}}+1}\right)}{{{\xi}_{0}}},\\
&& \mu_{3}=\pm\frac{\sqrt{3}\sqrt{6{{{\xi}_{0}}^{2}}+1}}{{{\xi}_{0}}}.
\eea 
From these eigenvalues we find that point $S_{2+}$ is an unstable node for $\gamma_{DE}<1+\sqrt{2}-\alpha$. Also, it is seen that $S_{2+}$ is a stable node for 
$\xi_{0}>\xi_{0}^{*}(\gamma_{DE},\alpha)$ with $1-\sqrt{2}-\alpha<\gamma_{DE}<1-\alpha$ and  $1-\sqrt{2}<\alpha<1+\sqrt{2}$ or for all values of $\xi_{0}$ and 
$1-\alpha<\gamma_{DE}<1$. This point is also a saddle point for  $\xi_{0}<\xi_{0}^{*}(\gamma_{DE})$ with  $-\sqrt{2}+1-\alpha<\gamma_{DE}<1-\alpha$ and $1-\sqrt{2}<\alpha<1+\sqrt{2}$ or  for all values of $\xi_{0}$ with  $\gamma_{DE}<-\alpha-\sqrt{2}+1$ 
and $1-\sqrt{2}<\alpha<1+\sqrt{2}$ or  for all values of $\xi_{0}$ with 
$\gamma_{DE}<1$ and $\alpha<1-\sqrt{2}$. Here, the function $\xi_{0}^{*}(\gamma_{DE},\alpha)$ is defined by
\be
\xi_{0}^{*}(\gamma_{DE},\alpha)=\frac{2 \left( {{\gamma}_{DE}}+\alpha-1\right) }{\sqrt{3}\left( {{{\gamma}_{DE}}^{2}}+2\alpha {{\gamma}_{DE}}-2 {{\gamma}_{DE}}+{{\alpha}^{2}}-2\alpha-1\right) }.
\ee
Finally, for the critical point $S_{3}$ we find the eigenvalues
\bea
&&{{\mu}_{1}}=\frac{\mathcal{E}-\sqrt{\mathcal{E}^2+9{{\xi}_{0}}{{C}^{3}}\left({{C}^{2}}-\frac{4}{3}\right)\mathcal{F}}}{4 {{\xi}_{0}}{{C}^{2}}},\\
&& {{\mu}_{2}}=\frac{\mathcal{E}+\sqrt{\mathcal{E}^2+9{{\xi}_{0}}{{C}^{3}}\left({{C}^{2}}-\frac{4}{3}\right)\mathcal{F}}}{4 {{\xi}_{0}}{{C}^{2}}},\\
&& \mu_{3}=3 {{\gamma}_{DE}}+3\alpha-4, 
\eea where we have defined 
\bea
&&\mathcal{E}=3 C^2\left[ {{\xi}_{0}}\left(2 {{\gamma}_{DE}}+ 3 \alpha-2\right)- {{C}}\right]-4{{\xi}_{0}}\alpha,\\
&&\mathcal{F}=3 C\gamma_{DE}\left[2 {{\xi}_{0}}\left( {{{\gamma}_{DE}}}+ \alpha-2\right)-3 {{C}}\right]+\nonumber\\
&& 9 {{C}^{2}}-6 \xi_{0}\left(\alpha+1\right) C-4\alpha, 
\eea and $\xi_{0}$, $\gamma_{DE}$, $\alpha$ and $C$ satisfy the relation \eqref{Alpha}.
Also, the determinant of the matrix of perturbations evaluated for this point is equal to
\be
\det(M)=-\frac{3\left( 3{{C}^{2}}-4\right) \left( 3{{\gamma}_{DE}}+3\alpha-4\right)\mathcal{F}}{16{{\xi}_{0}}C}.
\ee Since we have $C^2<4/3$, this critical point is stable for $\mathcal{E}<0$ and $\mathcal{F}>0$, with the condition $\gamma_{DE}<4/3-\alpha$. It is unstable for  $\mathcal{E}>0$ or  $\mathcal{F}<0$ or $\gamma_{DE}>4/3-\alpha$. For example, for $C=0.61$, $\gamma_{DE}=-0.4$, $\xi_{0}=0.01$ and $\alpha$ in the range $(0.33,0.39)$, we obtain that $S_{3}$ is a stable node, with $\mu_{1}<0$, $\mu_{2}<0$ and $\mu_{3}<0$.

\end{document}